\documentclass[reprint, aps, per,superscriptaddress]{revtex4-1}
\usepackage{hyperref}

\usepackage{graphicx}

\begin{document}

\title{Interpreting gains and losses in conceptual test using Item Response Theory}%

\author{Jean-Fran\c cois Parmentier}%
\email{jean-francois.parmentier@univ-tlse3.fr}
\affiliation{Universit\'e de Toulouse, UPS, IRES, F-31400 Toulouse, France}

\author{Brahim Lamine}%
\email{brahim.lamine@irap.omp.eu}
\affiliation{Universit\'e de Toulouse, UPS-OMP, CNRS, IRAP, F-31028 Toulouse, France}

\date{\today}%

\begin{abstract}
Conceptual tests are widely used by physics instructors to assess students' conceptual understanding and compare teaching methods. It is common to look at students' changes in their answers between a pre-test and a post-test to quantify a transition in student's conceptions. This is often done by looking at the proportion of incorrect answers in the pre-test that changes to correct answers in the post-test -- the gain -- and the proportion of correct answers that changes to incorrect answers -- the loss. By comparing theoretical predictions to experimental data on the Force Concept Inventory, we shown that Item Response Theory (IRT) is able to fairly well predict the observed gains and losses. We then use IRT to quantify the student's changes in a test-retest situation when no learning occurs and show that $i)$ up to 25\% of total answers can change due to the non-deterministic nature of student's answer and that $ii)$ gains and losses can go from 0\% to 100\%. Still using IRT, we highlight the conditions that must satisfy a test in order to minimize gains and losses when no learning occurs. Finally, recommandations on the interpretation of such pre/post-test progression with respect to the initial level of students are proposed.  \end{abstract}

\maketitle

\section{Introduction}
Conceptual tests are widely used by physics instructor to asses students' conceptual understanding and compare teaching methods. In particular, the Force Concept Inventory~\cite{hestenes_force_1992} (FCI) evaluate student's mastering of Newton laws~\cite{hake_interactive-engagement_1998}. It consists of 30 multiple-choice questions where incorrect answers are based on the most frequently answers given by students in interviews. Many topics are covered by the FCI : kinematics, identification of forces and the three Newton's laws~\cite{hestenes_force_1992, scott_exploratory_2012}. Instructors usually use the raw score or the Hake gain~\cite{hake_interactive-engagement_1998} to evaluate global student's progression. Item Response Theory (IRT) provide a more theoretically grounded measure of student's progression~\cite{wright_observations_1989, wright_history_1997, wallace_concept_2010}. Over the past decade, IRT have been applied with success to concept inventories, in particular to the FCI~\cite{morris_testing_2006, planinic_rasch_2010, wang_analyzing_2010, morris_item_2012, han_dividing_2015}. Student's raw score or student's proficiency given by IRT provide a global measure of the acquisition of the Newtonian concepts. 

A closer look to student's answer in a test-retest situation has shown that while the total score to the test is highly reliable, 31\% of the student's answers change from test to retest, suggesting weak reliability for individual answers~\cite{lasry_puzzling_2011}. Looking  how answers of students change between a pre-test -- before instruction -- and a post-test -- after instruction -- using a database embedding more than 13\,000 students' answers, Lasry et al.~\cite{lasry_two_2014} revealed a strong positive correlation between the initial score and the proportion of incorrect answers on the pre-test that were changed to correct answers on the post-test -- the gains. A symmetric result was found for the losses -- the proportion of correct answers on the pre-test that were changed to incorrect answers on the post-test, strongly and negatively correlated to the initial score. This result suggests that students with higher prior level learn more and forget less than students with lower prior level.

In this article we show that IRT can be used to qualitatively predict those experimental data while offering another interpretation of the previous results. The observed correlation mainly comes from inherent properties of the test rather than reflecting the level of progression of students. We show in particular that the student's proficiency progression, as obtained by IRT, increases for low proficiency students, a conclusion at the opposite of the previous interpretation.

The article is organized as follow : section~\ref{section:GL} provides definition of gains and losses; section~\ref{section:IRT} introduces IRT theory and the underlying assumptions; section~\ref{section:IRT_GL} compares theory's predictions with experimental data; section~\ref{section:IRT_changes} exploit IRT to predict answer's changes; finally section~\ref{section:GL_learning} and~\ref{section:conclusion} discuss and conclude this work.

\section{gains and losses}\label{section:GL}

\begin{figure}
  \includegraphics{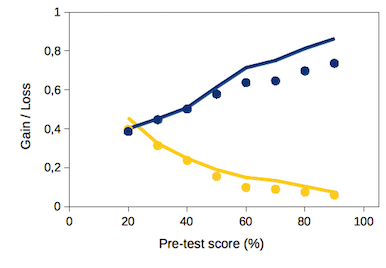}
  \caption{\label{fig:GLmazurIRT} 
  Gain (blue) and loss (yellow) as a function of pre-test score at the FCI. Points are measurements from a large pool of students~\cite{lasry_two_2014} and lines are theoretical predictions using questions parameters of the IRT analysis obtained in~\cite{wang_analyzing_2010}.}
\end{figure}

Consider the situation of students taking a same test two times : the first one before instruction and the second one after instruction. It is hoped that the score of each student increases, so that a part of answers which were initially wrong becomes correct. Following Lasry et al.~\cite{lasry_two_2014}, we define the gain $G$ as the proportion of incorrect answers on the pre-test that change to correct answers on the post-test. Similarly, the loss $L$ is defined as the proportion of correct answers on the pre-test that change to incorrect answers on the post-test. We then introduce $IC_i$ as the proportion of students who change from an incorrect ($I$) to a correct ($C$) answer at the question $i$ and $I_i$ as the proportion of initial incorrect answers. gains and losses are then defined by $G = \overline{IC_i} / \overline{I_i}$ and $L = \overline{CI_i} / \overline{C_i}$, where $\overline{( \ . \ )}$ denotes the average over the questions of the test. $C_i$ is the proportion of initial correct answers to question $i$ so that $\overline{C_i}$ is the average pre-test score of the students. Using data from more than 13,000 students' answers on the Force Concept Inventory (FCI), Lasry et al.~\cite{lasry_two_2014} measured dependance of gains and losses with prior knowledge (pre-test score). As shown in Fig.~\ref{fig:GLmazurIRT}, students with higher prior knowledge have higher gain and smaller loss than students with lower prior knowledge. In order to interpret these results, it is first necessary to draw the same graph when no learning occurs. That is to say when the same test is taken two times consecutively, with student not memorizing their previous answers and not having learned anything between the two tests. We show in the next sections how IRT is able to answer this question.

\section{The Item Response Theory}\label{section:IRT}

Item Response Theory (IRT) belongs to the family of latent trait modeling~\cite{kamata_note_2008}. In those models, each student is described by a number of latent traits, also call proficiencies. The answer of a student to a question is thought of as the result of the interaction between the capabilities of the person taking the test and the characteristics of the test items. The score of a student to an item is modeled by a probabilistic function of his proficiencies and the item's characteristics. A consequent number of knowledge and skills are always necessary to give a correct answer~\cite{reckase_multidimensional_2009} but in many cases, only one proficiency is sufficient to determine the student score. This is call unidimensional Item Response Theory but is often simply called IRT. This assumption was shown to be valid to model student's answer to the FCI~\cite{planinic_rasch_2010, wang_analyzing_2010} and will be assumed in the following.

\begin{figure}
  \includegraphics{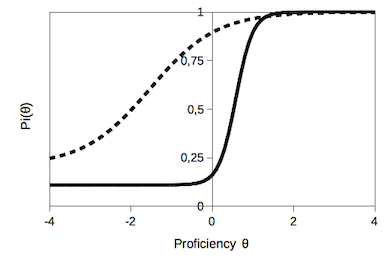}
  \caption{\label{fig:PiQ1Q13} Item characteristic curves for questions 1 (dashed line) and 13 (continuous line) of the FCI. Questions parameters are taken from~\cite{wang_analyzing_2010}.}
\end{figure}

Let's note $\theta$ the proficiency of a student. Each question $i$ is modeled by a function $P_i(\theta)$ which describes the probability of a student with proficiency $\theta$ to correctly answer to the question $i$. $P_i$ functions, called item characteristic curves, are often assumed to be generic "S-shape" functions (see Fig.~\ref{fig:PiQ1Q13}), called logistic function, whose variations characterize each questions. In the three-parameter item model, $P_i(\theta)$ is given by
\begin{equation}
\label{eq:Pi3PL}
P_i(\theta) = c_i + \frac{1-c_i}{1 + \exp\left[-1.7 \, a_i (\theta - b_i) \right] } \ ,
\end{equation}
where $a_i$, $b_i$ and $c_i$ are parameters of the question : $a_i$ is its discrimination power, $b_i$ its difficulty and $c_i$ the probability of guessing. The parameters are estimated by statistical techniques using a large pool of students answers. Other models exist such as the two-parameter model ($c_i=0$), the Rasch model ($c_i=0$ and $a_i=1$) and the non-parametric kernel smoothing approach~\cite{ramsay_kernel_1991}. All these models have been applied to the FCI~\cite{morris_testing_2006, planinic_rasch_2010, wang_analyzing_2010,morris_item_2012, han_dividing_2015}. For instance, $P_i$ functions for question 1 and 13 of the FCI are plotted in Fig.~\ref{fig:PiQ1Q13}. Question 13 is more difficult than question 1 ($b_{13} > b_1$) so its curve is more "on the right" of the graph. Its discrimination is also larger ($a_{13}>a_1$) so that the S-shape is steeper. Finally, the guessing parameter is lower ($c_{13} < c_1$), as seen on the value of $P_i$ when $\theta$ goes to $-\infty$.

The true score (in \%) of a group of students with proficiency $\theta$ is given by $S(\theta) = \overline{P_i(\theta)}$. Because of the probabilistic nature of IRT, the score $S(\theta)$ for a given proficiency $\theta$ differs from the observed score of a student with that proficiency $\theta$ -- the number of correct answer given by the student divided by the number of questions. The true score $S(\theta)$ is only recovered as an average over a large number of equal-proficiency student's individual observed scores. 
The observed score is also named the raw score and one strength of IRT is to convert this raw score, which is a discrete bounded variable, into a continuous unbounded variable, $\theta$, which is assumed to be an interval scale -- i.e. a scale which can be used to quantify a progression or a difference of proficiency between students \cite{wright_observations_1989}. 

\section{IRT prediction of gains and losses}\label{section:IRT_GL}

The objective of a course is to increase student's proficiency. Let's write $\theta_{pre}$ the proficiency of a student before instruction and $\theta_{post}$ its proficiency after instruction. By definition, the probability of choosing the correct answer to the question $i$ during the pre-test is $P_i(\theta_{pre})$. For the same reason, this probability is  $P_i(\theta_{post})$ for the post-test. For a wide group of student with the same proficiencies, we get $I_i = 1 - P_i(\theta_{pre})$ and $IC_i = (1-P_i(\theta_{pre})) \, P_i(\theta_{post})$. Reporting these equations into the definition of the gain and the loss leads to
\begin{equation}
\label{eq:G2}
G = S_{post} - \frac{\overline{ \delta P_i(\theta_{pre}) \, \delta P_i(\theta_{post}) } }{1 - S_{pre}} \ ,
\end{equation}
\begin{equation}
\label{eq:L2}
L = (1 - S_{post}) - \frac{\overline{ \delta P_i(\theta_{pre}) \, \delta P_i(\theta_{post}) } }{S_{pre}} \ ,
\end{equation}
where $\delta P_i(\theta)$ is the difference between probability of success of question $i$ and average test score $S$ for a given proficiency :
\begin{equation}
\delta P_i(\theta) = P_i(\theta) - \overline{P_i(\theta)} \textbf{.}
\end{equation}
By definition $\overline{\delta P_i(\theta)} = 0$. In the particular case when $\theta_{pre}=\theta_{post}$ (i.e. when no instruction occurs), $\overline{ \delta P_i \, \delta P_i }$ is the variance of the $P_i$'s for a given $\theta$ and is a characteristic of the test.

Equations~(\ref{eq:G2}) and (\ref{eq:L2}) show that IRT enables us to predict measured values for $G$ and $L$ once $\theta_{pre}$, $\theta_{post}$ and all the $P_i$'s are known. However, data of Lasry et al.~\cite{lasry_two_2014} give values of $G$ and $L$ as functions of $S_{pre}$ so informations about $\theta_{pre}$, $\theta_{post}$ and all the $P_i$'s function are missing. 

First $P_i$ functions are taken form literature. Using the three-parameter model, Wang and Bao~\cite{wang_analyzing_2010} performed an IRT analysis of the FCI using their own database of 2\,800 student's answers, leading to the knowledge of the 30~$P_i$ functions. The measurements obtained by Wang and Bao with their students can be used for any students because characteristics of questions are independent of the population used to obtained them. This property is known as parameter invariance~\cite{rupp_understanding_2006}. Hence there $P_i$ functions are used here.


Secondly, foreach values of $S_{pre}$ we estimated $S_{post}$ from data of Lasry et al.~\cite{lasry_two_2014} using
\begin{equation}
\label{eq:SpostSpreGL}
S_{post} = S_{pre} \, (1 - L) + (1 -  S_{pre}) \, G \ ,
\end{equation}
which comes from the definition of $G$ and $L$ and the fact that $S_{pre} = \overline{C_i}$. 

And finally $\theta_{pre}$ and $\theta_{post}$ are estimated by reversing the relation giving $S$ as a function of $\theta$ : $S(\theta) = \overline{P_i(\theta)}$. This is an approximation where the observed raw score is assumed to be equal to the true score. The sample of Lasry et al.~\cite{lasry_two_2014} contains $13\,000$ students divided into 9 bins leading to an average of $1\,400$ students for each raw score. In this case the hypothesis of equating the raw score to the true score seems reasonable.


Figure~\ref{fig:GLmazurIRT} shows that eqs.~(\ref{eq:G2}) and (\ref{eq:L2}) match fairly well the experimental measurements, indicating that IRT is able to correctly predict gains and losses. Discrepancies can be attributed to both uncertainties of measurements of $P_i$ and to an unperfect parameter invariance. Such a case can occur in particular when the hypothesis of unidimensionality does not hold. As shown by Scott and Schumayer~\cite{scott_exploratory_2012}, while a unique proficiency can be used to describe student's characteristic, a 5 dimensional model seems preferable. Our results show that a one-dimensional model is able to give the global tendency for the gain and the loss. A more detailed analysis is reported for future work.
%



\begin{figure}
  \includegraphics{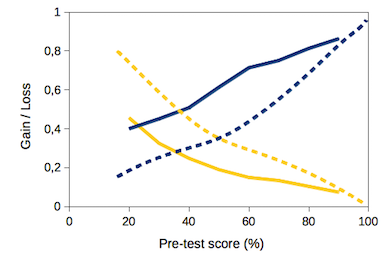}
  \caption{\label{fig:GLTRT} Gain (blue lines) and loss (yellow lines) as a function of pre-test score at the FCI. Continuous lines are IRT predictions when learning occurs, dashed lines are IRT predictions when no learning occurs (i.e. assuming $\theta_{post} = \theta_{pre}$).}
\end{figure}

As seen in Fig.~\ref{fig:GLmazurIRT}, gain is an increasing function of student's initial score. A tempting interpretation is to say that students with higher initial knowledge learn more than students with lower initial knowledge. The reverse is also true for loss : students with higher initial knowledge have lower loss than students with lower initial knowledge. However this argument implicitly assumes that gains and losses are zero when no learning occurs. We now show that this is not the case, which at least makes the previous conclusion unsecured. To do so, we use IRT to estimate $G$ and $L$ when $\theta_{post} = \theta_{pre}$, using equations~(\ref{eq:G2}) and (\ref{eq:L2}). Results are plotted in Fig.~\ref{fig:GLTRT}, which clearly show that even when no learning occurs gain is an increasing function of the pre-test score and raise up to one. Similarly, loss goes down from one to zero as pre-test score increases. For a pre-test score value of 50\% both gains and losses have the same value around 35\%. 
 Such a change in student answers at the same question has been observed between two successive passes of the FCI~\cite{lasry_puzzling_2011}. Reported values of gains and losses were 18\% and 20\% for a population mean score of 47\%. Discrepancy between their experimental measures and IRT prediction could largely be attributed to a memory effect because students took the tests two times in the same week so they may have memorized some of there initial answers. At the contrary, our IRT model assumes the independence between the test-retest, i.e. that students have not memorized any of their previous answers.

\section{Proportion of answer's change}\label{section:IRT_changes}

\begin{figure}
  \includegraphics{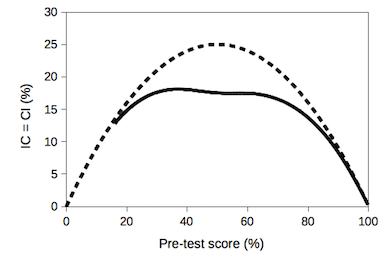}
  \caption{\label{fig:ICTRT} Proportion of answer's changes from a right (resp. wrong) answer to a wrong (resp. right) answer (continuous line) for the FCI. Dashed line is $S \, (1-S)$.}
\end{figure}

In order to interpret why gains and losses can have such high values even when no learning occurs, we focus directly on the global proportion of answer's change. In a test-retest situation, we have :
\begin{equation}
\label{eq:IC-TRT}
\overline{IC_i} = \overline{CI_i} = S \, (1-S) - \overline{\delta P_i^2} \ ,
\end{equation}
where $S=S_{pre}=S_{post}$. The explicit dependence of $S$ and $\delta P_i$ with $\theta_{pre} = \theta_{post}$ have been omitted for clarity. The first term of the right hand side of equation~(\ref{eq:IC-TRT}) is a parabolic function of $S$ and does not depend on the considered test. Hence, for any conceptual test, this part is identical. The second term on the right hand side of equation~(\ref{eq:IC-TRT}) depends on the item characteristic curves and consequently on the test. Values of $\overline{IC_i}$ have been plotted for the FCI as a function of the score in Fig.~\ref{fig:ICTRT}. It is clear that in this case, the contribution of $\overline{\delta P_i^2}$, while not negligible, is rather small. Consequently, for a group of students with a true score of 50 \%, nearly 18 \% of answers change from correct (resp. incorrect) to incorrect (resp. correct) in a test-retest situation.
This result has a consequence on the reliability of the test and on the interpretation of gains and losses. In order to interpret gains and losses in term of learning outcome, their values should be as small as possible in a test-retest situation. As a consequence, values of $\overline{IC_i}$ should also be as small as possible. Because the first term of equation~(\ref{eq:IC-TRT}) does not depend on the test, one can only influence the $\overline{\delta P_i^2}$ term in order to make it as high as possible (so that $\overline{IC_i}$ decreases). It immediately leads to the conclusion that one has to choose questions -- therefore the $P_i$'s functions -- in order to maximize values of $\overline{\delta P_i^2}$ for all $\theta$.

\begin{figure*}
  \includegraphics{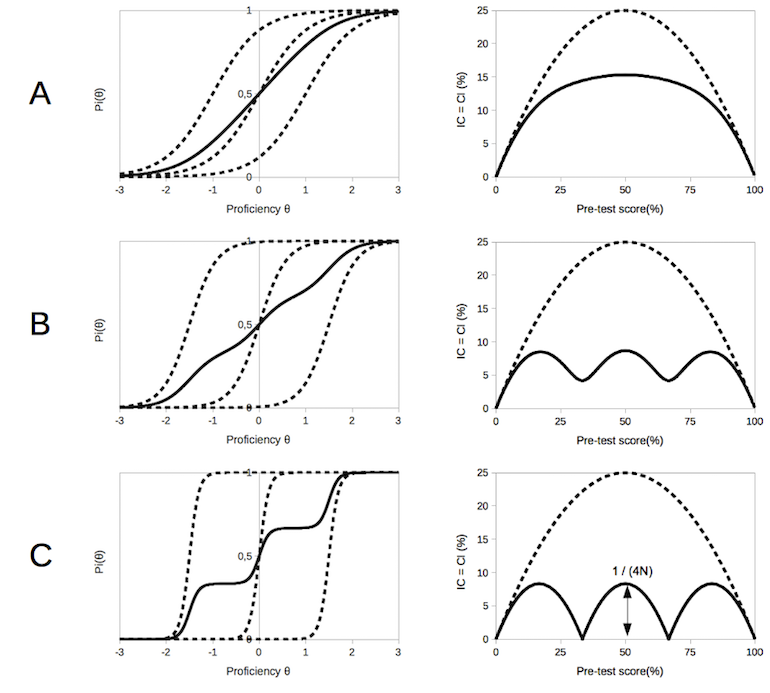}
  \caption{\label{fig:courbes3Tests} Each row corresponds to given tests (A, B or C) comprising 3 questions. Left : item characteristic curves of the three questions (dashed lines) and true score (continuous lines) as functions of proficiency $\theta$. Right : proportion of answer's change $\overline{IC_i} = \overline{CI_i}$ (continuous line) and $S(1-S)$ (dashed-line) as functions of the true score $S$. }
\end{figure*}

In order to understand how to choose those $P_i$'s functions, we consider the simple case of a test with only 3 questions. Three different cases are considered, each one corresponding to a particular set of $P_i$'s functions. The three cases are named test A, B and C and their item characteristic functions are plotted in Fig.~\ref{fig:courbes3Tests} (left column).
For each $\theta$, the proportion of answer's change is given by $\overline{CI_i} = \overline{P_i \, (1-P_i)}$, where $\overline{( \ . \ )}$ denotes the averaging over the 3 questions of the test. Hence, each individual question $i$ has a contribution of $P_i \, (1-P_i)$. This contribution is null when $P_i=0$ or $1$ and has a maximal value of $0.25$ when $P_i=0.5$. 


Test A has three questions whose characteristic curves overlap for a wide range of $\theta$. As a consequence, for a wide range of $\theta$ all individual questions will contribue to the proportion of answers that change. For instance, for a true score of 50\% ($\theta=0$), $P_1(\theta) = 0.88$, $P_2(\theta) = 0.5$, and $P_3(\theta) = 0.12$, leading to $P_1 \, (1-P_1) = P_3 \, (1-P_3) = 0.1$ and $P_2 \, (1-P_2)=0.25$. Hence, for a score of 50\%, the proportion of change, which is the average of these three values, is about 15\%. The representative curve of $\overline{IC_i}$ is very similar to the one obtained for the FCI, indicating that a lot of item characteristic curves of the FCI overlap, as already noted in previous studies analyzing the FCI using a unidimensional IRT~\cite{morris_testing_2006, planinic_rasch_2010, wang_analyzing_2010, morris_item_2012, han_dividing_2015}.

At the opposite, test C has three questions whose characteristic curves do not overlap - i.e. the range of $\theta$ where these functions go from a value close to 0 to a value close to 1 are well separated (see Fig.~\ref{fig:courbes3Tests}). As a consequence, each question will contribute separately to the proportion of answer's change. For instance, for a true score of 50\% ($\theta=0$), $P_1(\theta) \simeq 1$, $P_2(\theta) = 0.5$, and $P_3(\theta) \simeq 0$, leading to $P_1 \, (1-P_1) = P_3 \, (1-P_3) \simeq 0$ and $P_2 \, (1-P_2)=0.25$. Hence, for a score of 50\%, the proportion of change -- which is the average of the $P_i$ values -- is $0.25 / 3 \simeq 0.08$. This value is much smaller than for test A. In a test with $N$ separated questions, the maximal value of $\overline{IC_i}$ is $0.25/N$ and is obtained for values of $S = 0.5/N$, $1.5/N$, ... , $(N-0.5)/N$. In a test with $N=30$ separated-questions, maximal value for $\overline{IC_i}$ is about 1\%. Hence the change of answers occurs very rarely, and values of gains and losses remain very small.

Finally test B shows the transition between test A and the extreme case of test C.


\section{Interpretation of gains and losses when learning occurs}\label{section:GL_learning}

According to the discussion of the previous section, the interpretation of gains and losses should be separated in two extreme cases : when a wide majority of item characteristic curves overlap -- like in test A -- and when none of the item characteristic curves overlaps -- like in test C.

\begin{figure}
  \includegraphics{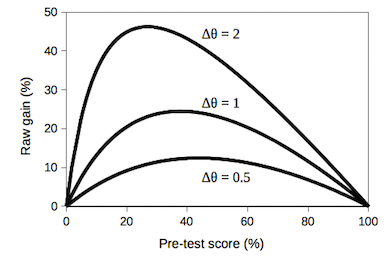}
  \caption{\label{fig:RawGainSpre} Evolution of the raw gain with initial pre-test score for three fixed values of student's learning $\Delta \theta$. The raw gain corresponds to $\Delta G$ for a type A test.}
\end{figure}

In the first case, $\overline{\delta P_i^2}$ is small and equations~(\ref{eq:G2}) and (\ref{eq:L2}) reduce to $G=S_{post}$ and $L=1-S_{post}$. Hence the gain is more or less the post-test score and does not add any supplementary informations on student's learning. One can still want to isolate the part of the gain due to instruction by defining $\Delta G = G_{\text{learning}} - G_{\text{no learning}}$. In the case of type A test, $\Delta G= S_{post}-S_{pre}=g_{raw}$, leading to the so-called raw gain (because $G=S_{pre}$ when no learning occurs). The analysis of Lasry et al.~\cite{lasry_two_2014} data shows that $g_{raw}$ is a decreasing function of the pre-test score. Does it mean that students with lower initial knowledge gain more than students with higher initial knowledge ? No because student's post score is limited to $100 \%$ so the raw gain $g_{raw}$ tends to zero when the pre-test score tends to $100 \%$. Also the score is an ordinal scale and not an interval scale~\cite{wright_observations_1989, wright_history_1997, wallace_concept_2010}. As a consequence, the raw score can only lead to a sorting of students but an increase of 1 point for a student with a low initial score does not reflect the same learning than an increase of 1 point for a student with a high initial score. A correct comparison of progress has to invoque an interval scale such as the student proficiency $\theta$ introduced in the previous sections~\cite{wright_observations_1989, wright_history_1997, wallace_concept_2010}. Fig.~\ref{fig:RawGainSpre} plots the raw gain as a function of the pre-test score for given values of student’s learning increase $\Delta \theta$. As seen on this figure, a given value of $g_{raw}$ corresponds to various value of student's progression $\Delta\theta$, depending of the initial student's score.

\begin{figure}
  \includegraphics{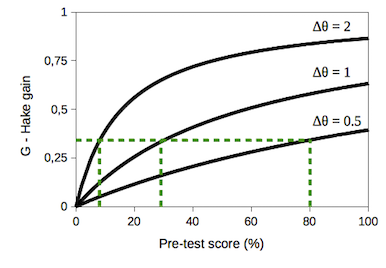}
  \caption{\label{fig:HakeGainSpre} Evolution of the Hake gain with initial pre-test score for three fixed values of student's learning $\Delta \theta$. Green dashed line is $G=0.34$ and correspond to $\Delta \theta=2$ for S=8\%, $\Delta \theta=1$ for S=30\% and $\Delta \theta=0.5$ for $S = 80\%$. The Hake gain corresponds to $\Delta G$ for a type C test.}
\end{figure}

\begin{figure}
  \includegraphics{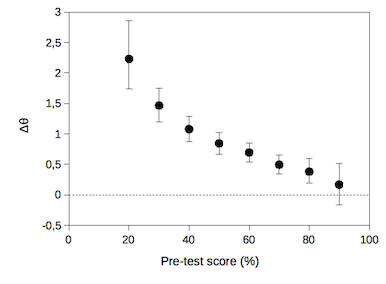}
  \caption{\label{fig:deltaThetaMazur} Evolution of student's learning ($\Delta \theta = \theta_{post} - \theta_{pre}$) with the pre-test score evaluated from data of Lasry et al.\cite{lasry_two_2014}.}
\end{figure}

In the second case (test of type C), where all questions are well separated, the proportion of questions that changes when no learning occurs is nearly null -- it is lower than 5\% for $N \ge 5$. Assuming a student positive progression $\Delta \theta = \theta_{post} - \theta_{pre}$ greater than the error range of all questions (i.e. $\forall i, \ \Delta \theta \gg 1/a_i$ with $a_i$ the discrimination power), the number of answers that change from incorrect to correct is $S_{post} - S_{pre}$ leading for the gain to
\begin{equation}
\label{eq:GintFb}
G \ = \ (S_{post} - S_{pre}) \ / \ (1-S_{pre})=G_{\text{Hake}}  \ .
\end{equation}
Interestingly, one recovers in this limit the Hake's gain~\cite{hake_interactive-engagement_1998}, which can be interpreted as the proportion of questions changing from incorrect to correct in a test comprising seperated item response curves (like test C). The number of answers that change from correct to incorrect is null and $L=0$. However, like the raw gain, the Hake gain is not an interval scale~\cite{wallace_concept_2010} and has to be taken with due care when comparing student's progression, as already emphasized. To illustrate this, let's consider an hypothetical test where the true score is a logistic function of the proficiency : $S = (1 + \exp(-\theta))^{-1}$. This model is characteristic of a test where question's difficulties are distributed over the proficiency scale following a gaussian law : there are few easy questions, few hard questions and a wide majority of questions with an intermediate level of difficulty. The Hake gain is plotted on Fig.~\ref{fig:HakeGainSpre} as a function of the pre-test score for various fixed value of student's learning $\Delta \theta$ that are typical of student's learning (see for instance Fig.~\ref{fig:deltaThetaMazur} for typical values of $\Delta\theta$ in a mechanic course). As can be seen, the gain is an increasing function of the pre-test score for a fixed value of student's learning. Hence, the fact that the gain is larger for initial high level students than for initial low level students does not necessarily reveal that the initial high level students have learned more. Moreover, a given value of $G$ corresponds to various value of student's progression $\Delta\theta$, depending of the initial student's score. As shown in Fig.~\ref{fig:HakeGainSpre}, a fixed value of the gain -- for instance 0.34 -- correspond to a strong learning for low pre-test score ($\Delta \theta=2$ for S=8\%), a medium learning for medium pre-test score ($\Delta \theta=1$ for S=30\%) and a low learning for high pre-test score ($\Delta \theta=0.5$ for $S = 80\%$). This clearly shows that the Hake gain should not be used to compare student's progression when they have different pre-test score, even in test of type C.
%

Table~\ref{table:ResumeGL} summarizes values of $G$ and $L$ for the two limit cases. As can be seen, $\Delta G$ reduces to the raw gain for type A tests and to the Hake gain for type C tests.

\begin{table}
\begin{tabular}{ c | c | c | c}
   Type of test & G & L & $\Delta G$ \\
\hline
   A & $S_{post}$ & $1-S_{post}$ & $g_{raw}$ \\
   C & $G_{Hake}$ & $0$ & $G_{Hake}$
 \end{tabular}
 \caption{Summary of gains and losses for the different types of test. $\Delta G=G_{\text{learning}}-G_{\text{no learning}}$ is the difference in gain between a situation when learning occurs and a situation when no learning occurs, that is to say the part of the gain which is due to learning.}
\label{table:ResumeGL}
\end{table}

We conclude this section by discussing the efficiency of instruction with respect to the initial level of the students. As already emphasized, the proficiency $\theta$ has good properties \cite{wright_observations_1989, wright_history_1997, wallace_concept_2010} and hence could be used to determine the learning $\Delta\theta$ of a student, $\Delta \theta = \theta_{post} - \theta_{pre}$. This increase of proficiency is plotted in Fig.~\ref{fig:deltaThetaMazur} as a function of the pre-test score for the data of Lasry et al.~\cite{lasry_two_2014}. We have evaluated $\theta$ using the scores by inverting the relation $S(\theta)$. According to Lasry et al.~\cite{lasry_two_2014}, uncertainties on pre-test scores, gains and losses are about 2\%, leading to uncertainties on the post-test score of the same order of magnitude. These uncertainties lead to uncertainties on the proficiencies, particularly for low or high scores due to the 'S' shape of the curve, and are represented in Fig.~\ref{fig:deltaThetaMazur}.
If $\theta$ is assumed to be the good scale for measuring the learning, Fig.~\ref{fig:deltaThetaMazur} clearly shows that learning decreases as the pre-test score increases. This is an opposite conclusion with the first interpretation of the evolution of gains and losses with pre-test score, but in accordance with the evolution of $g_{\text{raw}}$ with pre-test score. It seems to state that our teaching methods are more efficient on students with low prior knowledge. We recall that this result is based on data from more than 13,000 students who had taken the FCI at the beginning and at the end of an introductory physics course in a large variety of institutions: US high schools (10,007) , three Canadian two-year colleges (971), a US public university (1560) and three top-tier private universities (884)~\cite{lasry_two_2014}. Due to possible correlations between students' prior knowledge and student's institution, this could reflect a difference between institutions. But this also could mean that it is more difficult in an introductory physics course to give the same increase of learning to students with high prior level knowledge than to students with low prior level knowledge. This discussion is out of the scope of this article but in order to answer this question one would have to evaluate $\Delta \theta$ for each student in a group following the same course with the same teacher, plotting the same curve as in Fig.~\ref{fig:deltaThetaMazur} and finally perform a comparison across institutions.

\section{Conclusion}\label{section:conclusion}

We have shown that IRT is able to fairly well predict experimental measurements of gains and losses with the FCI when learning occurs. In addition, IRT shows that values of gains and losses for the FCI are rather high even when no learning occurs. The reason being that item characteristics curves overlap. All errors associated to individual questions contribute together to the probability of answer's change, leading to a difficult interpretation of gains and losses. In such a case the gain is more or less the post-test score and does not reveal that initial high level students have learned more that initial low level students.

In the case where item characteristic curves do not overlap, answer's changes are very low, the gain reduces to the Hake gain while the losses drop to zero. 

We have shown that the effect of instruction can be assessed by looking to the proficiency increase instead of looking to the gain increase. The proficiency increases more for low-level student (i.e. low pre-test score).

\section*{Acknowledgments}

This project was supported by the Initiative d'Excellence (IDEX) from the Universit\'e F\'ed\'erale Toulouse Midi-Pyr\'en\'ees.

\bibliography{refbiblio}

\end{document}